%% file: main.tex
\documentclass[runningheads]{llncs}
\usepackage[T1]{fontenc}
\usepackage{graphicx}
\usepackage{cite}
\usepackage{amsmath,amssymb,amsfonts}
\usepackage{algorithmic}
\usepackage{graphicx}
\usepackage{textcomp}
\usepackage{bmpsize}
\usepackage{xcolor}
\usepackage{lipsum}
\usepackage[colorlinks=true,urlcolor=black]{hyperref}
\def\BibTeX{{\rm B\kern-.05em{\sc i\kern-.025em b}\kern-.08em
    T\kern-.1667em\lower.7ex\hbox{E}\kern-.125emX}}
\usepackage{longtable}
\usepackage{booktabs}
\usepackage{tabularx}
\usepackage{multirow}
\usepackage{pifont}
\usepackage{makecell}
\usepackage{array}      % p{..}
\usepackage{rotating}   % sidewaystable

\begin{document}
%
% \title{Incident Response Without Experts: Usable Cybersecurity Actions for Smart Home Residents}
% \title{Actionable Cybersecurity Guidance for Smart Home Incident Response: An Exploratory Review of Governmental Sources}
% \title{Actionable Security Guidance for Smart Homes: An Exploratory Review of Governmental Sources}
\title{Cybersecurity Guidance for Smart Homes:\\ A Cross-National Review of Government Sources}

\titlerunning{Actionable Cybersecurity Guidance for Smart Homes}

\author{Victor Jüttner\inst{1,2} \and Erik Buchmann\inst{1,2}}
\authorrunning{V. Jüttner and E. Buchmann}
\institute{Center for Scalable Data Analytics and Artificial Intelligence (ScaDS.AI) \and
Leipzig University, Germany\\
\email{\{victor.juettner,erik.buchmann\}@uni-leipzig.de}}

\maketitle              % typeset the header of the contribution

\input{00-abstract}
\input{01-introduction}

\input{02-related}

\input{03-method}

\input{04-results}
\input{05-discussion}

\input{06-conclusion}
\input{07-acknowledgements}

\bibliographystyle{splncs04}
\bibliography{literature}

\input{08-appendix}

\end{document}

%% file: 00-abstract.tex
\begin{abstract}
Smart homes are increasingly targeted by cyberattacks, yet residents often lack guidance when incidents occur. Since affected residents are likely to seek help from trustworthy sources, this paper asks: What actionable cybersecurity guidance do governments provide to smart home users whose systems have been compromised? To answer this question, we conduct an exploratory, user-centered review of governmental cybersecurity guidance for smart homes across eleven countries to identify and characterize the types of guidance governments provide and to systematize their content. Using a standardized search and screening process, we derive three emergent clusters: incident reporting, general security recommendations, and incident response. Our findings show that governments provide abundant general security advice and accessible reporting channels, but structured incident response guidance tailored to smart homes is rare. Only two sources offer step-by-step recovery guidance for non-expert users, highlighting a gap between preventive advice and post-incident support.

\keywords{Smart Home \and Cybersecurity Advice \and Incident Response \and Government Policy \and Usable Security}

\end{abstract}

%% file: 01-introduction.tex
\section{Introduction}
\label{sec:intro}

Smart homes are increasingly exposed to cybersecurity incidents, yet residents often lack clear, actionable guidance on how to respond once a compromise has occurred. Domestic smart homes are typically operated by non-expert users who must make security decisions under uncertainty, time pressure, and limited technical insight. When incidents occur, affected residents are therefore likely to seek external guidance to understand what actions to take and how to recover their systems.

Governmental agencies represent an important source of such guidance. Official government resources are non-commercial and likely to be perceived as trustworthy, and they are frequently referenced in public cybersecurity awareness efforts such as Canada’s \textit{Get Cyber Safe}~\cite{getcybersafeTvSmartDevices}, France’s \textit{cybermalveillance}~\cite{cybermalveillanceIot}, Japan’s \textit{NOTICE}~\cite{noticeGoJp}, or Australia’s \textit{Protect Yourself}~\cite{acscAdvSteps}. At the same time, it remains unclear what actionable guidance governments actually provide to smart home users following a cybersecurity incident, how this guidance is structured, and whether it supports users beyond general preventive advice.

\textbf{What actionable cybersecurity guidance do governments provide to smart home users whose systems have been compromised?} This paper addresses this gap by empirically examining governmental cybersecurity guidance for smart home users. We adopt the perspective of a smart home resident seeking help after an incident and focus on official governmental online resources that such users are likely to encounter.

To answer this question, we conduct an exploratory, user-centered review of governmental cybersecurity resources across eleven countries. Our goal is to identify and characterize the types of guidance governments provide and to systematize their content. To this end, we use a standardized search approach that reflects how affected residents might seek help to identify relevant official sources. All identified materials are manually screened and inductively clustered based on their primary type of guidance. This process leads to three categories: incident reporting, describing how users can notify authorities or relevant agencies; general security recommendations, focusing on preventive and hardening practices; and incident response, providing guidance on recovery after a compromise. We then systematize and analyze the content of each cluster.

Our analysis shows that governmental guidance for smart home users is both abundant and unevenly distributed. Across all countries, we find accessible channels for incident reporting and a wide range of general security recommendations, with substantial cross-agency agreement on baseline practices such as securing routers, updating devices, and strengthening authentication. In contrast, structured incident response guidance tailored to smart home environments is rare. Only two sources in our dataset provide step-by-step recovery guidance aimed at non-expert users.

By systematizing existing governmental recommendations and identifying where guidance is lacking, this paper contributes an empirical overview of official smart home cybersecurity advice and highlights the need for user-centered incident response frameworks for domestic environments.

%% file: 02-related.tex
\section{Related Work}
\label{sec:related}

Related work addresses smart-home security constraints, online cybersecurity advice for end users, and government guidance for consumer IoT security. 

\subsection{Smart home Security in Practice}
Smart-home security is shaped by home-specific constraints such as limited expertise, time, and tooling. Nthala and Flechais argue that home network security must be conceptualized around households’ lived realities; consequently, guidance must be feasible for non-experts and robust to uncertainty~\cite{nthala2018rethinking}. Ye et al.\ show that home routers ship with heterogeneous and sometimes risky default settings, and that resulting exposures can persist because users rarely revisit configuration unless prompted by an incident~\cite{ye2024exposed}. These constraints show that smart home security needs clear, low-effort guidance that residents can execute under stress~\cite{juettner2025actionable_notifications}.

% In reasearch this problem has been tackled by developing automatic technical solutions for detecting cyber attacks~\cite{sikder2021aegis, hub2027, ghost} and notifying users with actionable guidance for resolving incidents~\cite{sok-11-no-name, juettner2024chatids, juettner2025actionable_notifications}. 

\subsection{Online Cybersecurity Advice for End Users}
Redmiles et al.\ show that end users learn security practices from heterogeneous sources, and that advice sources are associated with whether recommended behaviors are adopted~\cite{redmiles2016learned}. The quality of sources also varies substantially: Redmiles et al.\ also highlight the differences of clarity and actionability of security advice on the web and show that end-users are left on their own to prioritize which actions they need to take~\cite{redmiles2020quality}. Quinlan et al.\ have shown an increase in vocabulary complexity and reading difficulty of security advice, which limits their practicability, especially for vulnerable users~\cite{quinlan2025efficacy}. Focusing on smart-home contexts, Turner et al.\ show that searching the web yields a broad mix of sources with differing credibility and device-specific relevance, making it difficult for users to identify appropriate guidance when issues arise~\cite{turner2021googling}. Taken together, the online security advice is fragmented: guidance varies in quality and specificity, and users are frequently left to judge credibility and relevance and to translate generic recommendations into context-appropriate actions.

\subsection{Government guidance for smart home security}
Governments bodies promote baseline consumer IoT security via awareness campaigns, public guidance portals, regulation, and labeling. Analyses of government-led cybersecurity awareness materials suggest that many initiatives rely heavily on education and assume that knowledge translates into behavior change~\cite{vansteen2020bct}; ENISA likewise highlights the centrality of awareness in national strategies while noting persistent challenges in targeting and  impact~\cite{enisa2021raising}. Rama and Keevy compare good practices across government-led cybersecurity awareness websites, illustrating variation in topic coverage and presentation~\cite{rama2023public}. These initiatives provide smart security hygiene and inform purchasing.

\subsection{Research gap}
Prior work shows that smart home security advice is widely available online, but varies substantially in clarity and actionability, often leaving users to judge relevance and translate generic recommendations into context-appropriate actions~\cite{redmiles2016learned,redmiles2020quality,quinlan2025efficacy,turner2021googling}. Governments represent a particularly important source of such advice, as their portals are likely to be perceived as trustworthy and commonly promote baseline consumer IoT security practices through public awareness and guidance materials~\cite{vansteen2020bct,enisa2021raising,rama2023public}.

Despite this, existing literature provides limited empirical insight into the content and structure of official governmental guidance for smart home users. In particular, it remains unclear what types of actionable guidance governments make available to users following a cybersecurity incident and how this guidance is organized across different contexts. Addressing this gap requires a systematic examination of governmental online resources beyond individual campaigns or isolated examples.

% Current research already shows how security advice for the smart home should be designed like and it shows that there is a  plethora of sources for security advice for the smart home available online. One source of security advice are official government websites, that are very likely to be trustworthy, they provide security advice for consumer smart home security hygiene. A research gap that still exist now is not only who or how security advice should be given but also what exactly the content of the advice is that is given to smart home user. For this purpose we analyze what governments are recommending that smart home end user do in case of and incident, we examine what categories of advice governments are providing in general and what they specifically recommend in case of an incident.

%% file: 03-method.tex
\section{Review Procedure}
\label{sec:method}

Our review procedure consists of four stages: purposive country selection, a standardized user-centered search for official sources, manual screening for relevance, and qualitative clustering of the identified sources.

\begin{figure}[h]
    \centering
    \includegraphics[width=\textwidth]{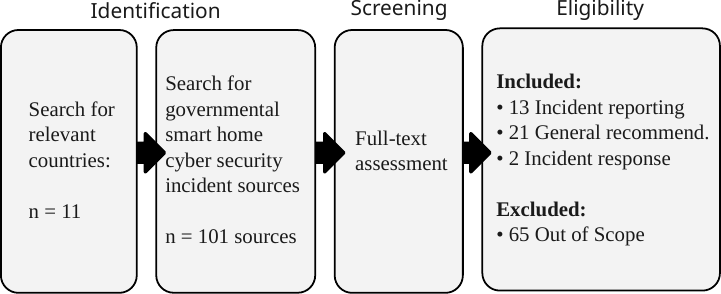}
    \caption{Data Collection}
    \label{fig:prisma}
\end{figure}

\subsection{Country Selection}
We employed purposive sampling to select countries for an exploratory, cross-national review of governmental smart home cybersecurity guidance. To balance cross-national coverage with feasibility, we targeted a sample size of approximately 10–13 countries. We considered countries eligible for inclusion if national authorities provided publicly accessible cybersecurity guidance for private individuals. We identified candidate countries through an exploratory search targeting official government sources. To avoid restricting the analysis to a single market or regional context, we further aimed to include countries that varied in consumer smart home adoption levels and geographic location.

We assessed differences in consumer smart home adoption using available international statistics and national surveys~\cite{OECD2023MeasuringIoT, CSA2023CybersecurityAwarenessSurvey2022_AnnexA}. These data confirmed that the selected countries span different levels of market maturity rather than uniformly high adoption. We achieved geographic diversity by including countries from North America, Europe, and the Asia--Pacific region, enabling comparison across different regulatory and governance contexts. Based on these criteria, we included eleven countries in the analysis: Australia, Austria, Canada, Finland, France, Germany, Japan, New Zealand, Singapore, the United Kingdom, and the United States.

\subsection{Data Collection}

For each selected country, we conducted a standardized, user-centered exploratory search. We used OpenAI’s ChatGPT (gpt-5.2-2025-12-11) via the WebUI as a search engine. For each country, we issued the prompt \textit{``My smart home devices have been hacked. I live in \{COUNTRY\}. Could you please provide me with official sources that offer advice on what I should do now?''}. This prompt reflects how a smart home resident might seek advice after an incident.

From ChatGPT’s responses, we collected only the referenced sources; we did not analyze the generated text itself. To ensure consistency across countries, we started a new chat session for each search. When ChatGPT directed us to a page that did not contain guidance but linked to another page that did, we included the linked page. We translated sources that were not available in English using the built-in translation function of the Firefox browser.

We initiated the search process on 29 December 2025. Across the eleven countries, we identified a total of 101 distinct websites. These websites originated from 49 distinct governmental institutions, including national cybersecurity agencies, consumer protection authorities, and police.

\subsection{Source Screening}

We manually verified all identified sources to ensure relevance and consistency with the scope of this study. We included sources only if they met all of the following criteria:
\begin{enumerate}
    \item The source originates from an official national authority.
    \item The source explicitly targets private individuals or households.
    \item The source offers actionable guidance, such as concrete steps for prevention, mitigation, or incident reporting.
    \item The source addresses smart homes, IoT in domestic settings, or home network security.
\end{enumerate}
We excluded sources that did not meet all criteria from further analysis.

\subsection{Clustering}

After screening, we analyzed the sources inductively to identify recurring patterns in their primary purpose and type of actionable guidance. We treated each source as a unit of analysis and assigned it to a cluster based on its dominant user-facing function (e.g., reporting, preventive guidance, or recovery support). We refined clusters iteratively as we reviewed additional sources, until no new categories emerged. After clustering, we analyzed each cluster using a qualitative content analysis approach tailored to the content type and scope of the respective cluster.

%% file: 04-results.tex
\section{Systematization}

After screening, we included 35 sources in the final dataset (Table~\ref{tab:source-types-country}) and excluded the remaining sources as out of scope. We qualitatively analyzed the included sources and grouped them into three clusters based on their primary type of guidance: incident reporting, general security recommendations, and incident response.

\begin{table}[h]
\centering
\caption{Distribution of guidance clusters by country}
\label{tab:source-types-country}
\begin{tabular}{lccc}
\toprule
Country & Reporting & Recommendations & Response \\
\midrule
Australia     & 2 & 4 & 0 \\
Austria       & 1 & 2 & 0 \\
Canada        & 2 & 3 & 0 \\
Finland       & 2 & 1 & 0 \\
France        & 0 & 2 & 1 \\
Germany       & 0 & 1 & 0 \\
Japan         & 0 & 1 & 0 \\
New Zealand   & 2 & 1 & 0 \\
Singapore     & 1 & 1 & 1 \\
UK            & 1 & 1 & 0 \\
USA           & 1 & 4 & 0 \\
\midrule
Total         & 12 & 21 & 2 \\
\bottomrule
\end{tabular}
\end{table}

We then analyzed each cluster using a qualitative content analysis approach tailored to its scope. For incident reporting sources, we examined the reporting channels offered (e.g., online forms, telephone hotlines, and email) and how sources framed incidents. For incident response sources, we analyzed the presence and structure of recovery guidance. For general security recommendation sources, we synthesized actionable steps mentioned across multiple sources and iteratively grouped them by their primary target, such as home routers, smart devices, online accounts, or general practices.

\subsection{Incident Reporting}

In our dataset, we identified 13 incident-reporting websites spanning 9 countries and operated by 11 distinct agencies. Across this incident-reporting cluster, agencies consistently offer a limited set of recurring reporting channels (Table~\ref{tab:incident-reporting-channels}). These include guided online reporting tools, dedicated telephone hotlines, and email-based reporting. In addition, several agencies provide referral pages that help users identify and navigate to the most appropriate reporting destination.

Across the incident-reporting sources, reporting is generally framed broadly as reporting cybercrime or cybersecurity incidents affecting individuals and small businesses. The sources accept reports of common compromise scenarios (e.g., hacking/account compromise, scams/fraud, malicious software). We did not find any smart home or home network specific incident reporting website.

\begin{longtable}{@{}p{4.3cm}*{11}{c}c@{}}
\caption{Incident reporting channels and which public agencies provide them.}
\label{tab:incident-reporting-channels}\\

\toprule
Reporting channel
& \rotatebox{90}{ACSC (AU)\cite{acsc_au_recover_hacking,acsc_au_report}}
& \rotatebox{90}{BKA (AT)\cite{bka_at_internetkriminalitaet}}
& \rotatebox{90}{CCCS (CA)\cite{cybercentre_ca_report_individuals}}
& \rotatebox{90}{RCMP (CA)\cite{rcmp_ca_report_cybercrime_fraud}}
& \rotatebox{90}{NCSC-FI (FI)\cite{traficom_fi_what_to_do_incident,traficom_fi_report_incident}}
& \rotatebox{90}{MBIE (NZ)\cite{mbie_nz_if_hacked}}
& \rotatebox{90}{NCSC-NZ (NZ)\cite{ncsc_nz_report_it}}
& \rotatebox{90}{CSA (SG)\cite{csa_sg_cyber_aid}}
& \rotatebox{90}{CoLP (UK)\cite{colp_uk_report_fraud_guide}}
& \rotatebox{90}{IC3 (US)\cite{fbi_us_ic3_complaint}}
& \rotatebox{90}{Total} \\
\midrule
\endfirsthead

\toprule
Reporting channel
& \rotatebox{90}{ACSC (AU)\cite{acsc_au_recover_hacking,acsc_au_report}}
& \rotatebox{90}{BKA (AT)\cite{bka_at_internetkriminalitaet}}
& \rotatebox{90}{CCCS (CA)\cite{cybercentre_ca_report_individuals}}
& \rotatebox{90}{RCMP (CA)\cite{rcmp_ca_report_cybercrime_fraud}}
& \rotatebox{90}{NCSC-FI(FI)\cite{traficom_fi_what_to_do_incident,traficom_fi_report_incident}}
& \rotatebox{90}{MBIE (NZ)\cite{mbie_nz_if_hacked}}
& \rotatebox{90}{NCSC-NZ (NZ)\cite{ncsc_nz_report_it}}
& \rotatebox{90}{CSA (SG)\cite{csa_sg_cyber_aid}}
& \rotatebox{90}{CoLP (UK)\cite{colp_uk_report_fraud_guide}}
& \rotatebox{90}{IC3 (US)\cite{fbi_us_ic3_complaint}}
& \rotatebox{90}{Total} \\
\midrule
\endhead

\midrule
\multicolumn{13}{r}{\emph{Continued on next page}}\\
\midrule
\endfoot

\bottomrule
\endlastfoot

Online reporting tool
    &2x &   & x &   &2x &      & x &   &   & x & 7 \\

Telephone reporting
    &2x &   & x &   &2x &      & x &   &   &   & 4 \\

Email reporting
    &   & x &   &   &   & x    & x &   &   &   & 3 \\

Referral page 
    & x &   & x & x &   &      &   & x & x &   & 5 \\

\end{longtable}

\subsection{Incident Response}

Two sources went beyond general cybersecurity recommendations by providing explicit guidance on recovery following a compromise: GIP ACYMA from France~\cite{cybermalveillanceIR} and the CSA in Singapore~\cite{csa2024iotdevices}. As only two sources in our dataset offer incident response guidance specifically for smart home environments, we do not systematize them further but instead present them individually.

GIP ACYMA provides a 12-step recovery plan that is structurally similar to industry incident response playbooks, but is tailored to non-expert users. It offers step-by-step instructions for mitigating an ongoing attack and re-securing the affected device or network. The guidance is intentionally generic and applies to a broad range of consumer systems, including computers, tablets, mobile phones, connected objects, servers, and home networks.

The CSA, by contrast, presents an extensive set of general cybersecurity recommendations alongside three recovery steps specifically targeted at smart home devices: \textit{Disconnect the Device from the Internet}, \textit{Change Credentials and/or Perform a Factory Reset}, and \textit{Contact the Manufacturer for Assistance}.

\subsection{General Recommendations}

For each of the eleven countries we could find at least one source that provided general cyber security recommendations to use smart home devices, in total we could identify 21 sources providing 46 distinct recommendations, provided by 17 government agencies~\ref{tab:agencies-all}. Our results show that recommendations are device specific, they are aimed at the router, smart device, online-account, or general recommendations. Most recommendation we found are either aimed at router, smart device or both.
% List single recommendations
All recommendations in Table~\ref{tab:cyber-recommendations-agencies} were referenced at least twice among all sources. Recommendations that were only mentioned once are reported for completeness in Table~\ref{tab:single_recs_nointerp} in the Appendix.

Several recommendations show broad cross-agency consensus, suggesting a widely accepted ``baseline'' for consumer cybersecurity. Across device categories, guidance most consistently emphasizes strengthening authentication (e.g., changing default credentials and using strong passwords), keeping software updated (including enabling automatic updates where available), and reducing attack surface and lateral movement risk (e.g., disabling unnecessary features and isolating smart devices via guest networks/VLANs). In contrast, a long tail of recommendations appears in only a small number of sources, indicating fragmentation or context-specific emphasis (e.g., monitoring internet usage, rebooting regularly, or advice about specific interfaces/communications).

\begin{longtable}{@{}p{4.3cm}*{16}{c}c@{}}
\caption{General recommendations by public agencies.}
\label{tab:cyber-recommendations-agencies}\\

\toprule
Recommendation
& \rotatebox{90}{ACSC (AU)\cite{acscAdvSteps,acscIotDevices,acscWifiRouter,acscIotTipsPdf}}
& \rotatebox{90}{A-SIT (AT)\cite{asitSmartHome,asitSmartHomeSystems}}
& \rotatebox{90}{CCCS (CA)\cite{cccsIotSecurityItsap00012}}
& \rotatebox{90}{GCS (CA)\cite{getcybersafeTvSmartDevices}}
& \rotatebox{90}{OPC (CA)\cite{opcSmartDevicesPrivacy}}
& \rotatebox{90}{NCSC-FI (FI)\cite{ncscfiHomeNetworkRouter}}
& \rotatebox{90}{GIP ACYMA (FR)\cite{cybermalveillanceIot}}
& \rotatebox{90}{MEFSIN (FR)\cite{minecofinConnectedObjectsRisks}}
& \rotatebox{90}{BSI (DE)\cite{bsiSmarthome}}
& \rotatebox{90}{NOTICE (JP)\cite{noticeGoJp}}
& \rotatebox{90}{CERT NZ (NZ)\cite{certnzUseIotSecurely}}
% & \rotatebox{90}{CSA (SG)\cite{csaSgProtectIot}}
& \rotatebox{90}{SPF (SG)\cite{spfSmartHomeCameras}}
& \rotatebox{90}{NCSC-UK (UK)\cite{ncscUkSmartDevices}}
& \rotatebox{90}{CISA (US)\cite{cisaHomeNetworkSecurity,cisaSecuringIot}}
& \rotatebox{90}{FTC (US)\cite{ftcSecuringInternetConnectedDevices}}
& \rotatebox{90}{IC3 (US)\cite{ic3Psa250605}}
& \rotatebox{90}{Total} \\
\midrule
\endfirsthead

\toprule
Recommendation
& \rotatebox{90}{ACSC (AU)\cite{acscAdvSteps,acscIotDevices,acscWifiRouter,acscIotTipsPdf}}
& \rotatebox{90}{A-SIT (AT)\cite{asitSmartHome,asitSmartHomeSystems}}
& \rotatebox{90}{CCCS (CA)\cite{cccsIotSecurityItsap00012}}
& \rotatebox{90}{GCS (CA)\cite{getcybersafeTvSmartDevices}}
& \rotatebox{90}{OPC (CA)\cite{opcSmartDevicesPrivacy}}
& \rotatebox{90}{NCSC-FI (FI)\cite{ncscfiHomeNetworkRouter}}
& \rotatebox{90}{GIP ACYMA (FR)\cite{cybermalveillanceIot}}
& \rotatebox{90}{MEFSIN (FR)\cite{minecofinConnectedObjectsRisks}}
& \rotatebox{90}{BSI (DE)\cite{bsiSmarthome}}
& \rotatebox{90}{NOTICE (JP)\cite{noticeGoJp}}
& \rotatebox{90}{CERT NZ (NZ)\cite{certnzUseIotSecurely}}
% & \rotatebox{90}{CSA (SG)\cite{csaSgProtectIot}}
& \rotatebox{90}{SPF (SG)\cite{spfSmartHomeCameras}}
& \rotatebox{90}{NCSC-UK (UK)\cite{ncscUkSmartDevices}}
& \rotatebox{90}{CISA (US)\cite{cisaHomeNetworkSecurity,cisaSecuringIot}}
& \rotatebox{90}{FTC (US)\cite{ftcSecuringInternetConnectedDevices}}
& \rotatebox{90}{IC3 (US)\cite{ic3Psa250605}}
& \rotatebox{90}{Total} \\
\midrule
\endhead

\midrule
\multicolumn{18}{r}{\emph{Continued on next page}}\\
\midrule
\endfoot

\bottomrule
\endlastfoot

\multicolumn{18}{@{}l}{\textbf{Router}}\\
Change admin credentials          & 3x & x &   &   & x & x &   & x & x &   &   &    x &   & x & x &   & 11 \\
Change SSID+Wi-Fi passw.          & 4x & x &   & x & x & x & x & x &   &   &   &      &   & x & x &   & 12 \\
Update regularly                  & 2x & x &   &   &   & x &   &   &   &   &   &      &   & x & x & x & 7 \\
Enable automatic updates          &   & 2x &   & x & x & x &   &   & x &   & x &      & x &   &   &   & 8 \\
Disable remote management         & 2x &   & x &   &   & x &   &   & x &   &   &      &   & x &   &   & 6 \\
Replace after EOL                 & 2x &   &   &   &   &   &   &   &   & x &   &      &   &   &   &   & 3 \\
Enable firewall                   &   &   &   &   &   & x &   &   & x &   &   &      &   & x &   &   & 3 \\
Use guest Wi-Fi                   & 3x & x & x & x & x & x & x & x & x &   & x &    x &   &   &   &   & 13 \\
Use WPA2/WPA3                     & 3x & x &   &   & x & x & x &   &   &   &   &    x &   & x & x &   & 10 \\
Check connected devices           & x &   &   &   &   &   &   &   &   &   &   &      &   & x &   &   & 2 \\
Monitor internet usage            & 3x &   &   &   &   &   &   &   &   &   &   &     &   &   &   & x & 4 \\
Access network over VPN           &   & x &   &   &   &   &   &   & x &   &   &      &   &   &   &   & 2 \\

\midrule
\multicolumn{18}{@{}l}{\textbf{Smart device}}\\
Change default credentials        & 3x & x & x & x & x &   & x &   & x & x & x &   x & x & x & x &   & 15 \\
Update regularly                  & 3x & x & x & x & x &   & x & x & x & x & x &    x & x & 2x & x & x & 18 \\
Enable automatic updates          &   & 2x &   & x & x & x &   &   & x &   & x &      & x &   &   &   & 8 \\
Disable remote management         & x & x &   &   & x &   &   &   & x &   &   &      & x &   &   &   & 5 \\
Replace after EOL                 &   &   &   &   &   &   &   &   & x & x & x &      &   &   &   &   & 3 \\
Check reviews / ratings           & 3x &   & x &   & x &   & x & x & x &   & x &    x & x &   &   &   & 11 \\
Avoid unnecessary internet        & 3x &   &   &   & x &   &   &   & x &   & x &      &   &   & x &   & 7 \\
Disable unused features           & 2x & x &   & x & x &   &   &   &   & x &   &    x &   & x & x &   & 9 \\
Turn off when not in use          & 3x & x &   & x & x &   & x &   &   &   &   &      &   &   &   &   & 7 \\
Reboot regularly                  & 3x &   &   &   &   &   &   &   &   &   &   &      &   &   &   &   & 3 \\
Secure physical access            & 3x &   &   &   &   &   &   &   & x &   &   &      &   &   &   &   & 4 \\
Encrypted communication           &   & x & x &   &   &   &   &   & x &   &   &      &   &   &   &   & 3 \\
Avoid insecure interfaces         &   &   &   &   & x &   &   &   &   &   &   &      &   & 2x & x & x & 5 \\

\midrule
\multicolumn{18}{@{}l}{\textbf{Online}}\\
Limit social media sharing        &   & x &   & x &   &   &   &   &   &   &   &      &   &   &   &   & 2 \\
Enable MFA                        & x & x & x &   &   &   &   &   & x &   &   &    x & x &   & x &   & 7 \\
Protect digital identity          &   & x &   &   & x &   & x &   &   &   &   &      &   &   & x &   & 4 \\

\midrule
\multicolumn{18}{@{}l}{\textbf{General}}\\
% Keep phone apps updated           &   &   &   &   &   &   & x &   &   &   &   &      &   &   &   &   & 1 \\
Use strong passwords              & x & x & x & x & x &   & x & x & x & x &   &    x & x & 2x &   &   & 13 \\
Use a password manager            & x & x &   &   &   &   &   &   & x &   &   &      &   &   &   &   & 3 \\
\end{longtable}

%% file: 05-discussion.tex
\section{Discussion}
\label{sec:discussion}

Our findings reveal a clear imbalance in the type of cybersecurity guidance governments provide to smart home users. While general security recommendations and incident reporting mechanisms are widely available and relatively consistent across countries, structured incident response guidance tailored to smart home environments remains rare. 
\subsection{Implications}

\paragraph{I1: General Recommendations}
The abundance of general recommendations indicates an emerging consensus on baseline smart home security practices, such as changing default credentials, keeping devices updated, and segmenting home networks. However, frequent mention of certain practices should not be interpreted as evidence that these measures are sufficient, optimal, or appropriate in all incident scenarios.

\paragraph{I2: Incident response}
The lack of structured incident response guidance is perhaps unsurprising. Unlike organizational IT environments, smart homes lack established response frameworks or reliable ways to assess the security status. Without such foundations, it is difficult for authorities to prescribe recovery steps that are both actionable and verifiable for non-expert users. As a result, most guidance stops short of helping users understand when an incident has been resolved.

\paragraph{I3: Incident Reporting}
At the same time, the ease of accessing reporting channels across countries is a positive finding. Governments have largely succeeded in making incident reporting visible and accessible, even if reporting is framed broadly as cybercrime rather than smart home–specific compromise.

\subsection{Design and Practice Recommenations}

\paragraph{R1: Step-by-Step Workflows}
Existing recommendations should be reorganized into structured, step-by-step workflows that users can follow under stress. Rather than presenting long lists of preventive measures, guidance could be ordered into phases (e.g., containment, remediation, hardening), allowing users to progress incrementally from immediate response to longer-term security improvements.

\paragraph{R2: Validation Mechanisms}
Second, current guidance lacks mechanisms for validation. Users are rarely given ways to assess whether their smart home is secure again after taking recommended actions. Designing lightweight validation cues, such as checks for unknown devices, confirmation of update status, or indicators of restored normal behavior could significantly improve user confidence and reduce premature termination of recovery efforts.

For practitioners and policymakers, this suggests that improving smart home incident response does not necessarily require inventing new recommendations, but rather systematizing existing advice and presenting it in a form that supports decision-making and reassurance.

\subsection{Limitations}

This study is exploratory and subject to several limitations. Most notably, source identification relied on a ChatGPT-assisted, web-based search strategy rather than a predefined registry of governmental resources. As a result, it is possible that relevant agencies, countries, or guidance documents were not identified. The country sample was purposive rather than exhaustive, and the findings do not claim global representativeness.

In addition, the analysis focused on the presence and structure of guidance rather than its real-world effectiveness. We did not evaluate whether following the identified recommendations would successfully resolve incidents, nor how users interpret or apply this guidance in practice.

%% file: 06-conclusion.tex
\section{Conclusion}
\label{sec:conclusion}

Smart home residents increasingly face cybersecurity incidents but often lack clear, actionable guidance tailored to domestic environments. In this exploratory review, we examined governmental cybersecurity resources across eleven countries to assess what official guidance is available to users whose smart homes have been compromised. While governments widely provide general security recommendations and accessible incident reporting channels, structured incident response guidance specific to smart home contexts remains rare.

By systematically characterizing and synthesizing existing governmental guidance, this work highlights a gap between preventive advice and post-incident support for non-expert users. Rather than proposing new response frameworks, our findings provide an empirical foundation for future efforts aimed at translating existing advice into clearer, more structured recovery guidance. Strengthening post-incident support has the potential to better equip smart home residents to navigate cybersecurity incidents and assess recovery in domestic settings.

%% file: 07-acknowledgements.tex
\begin{credits}
\subsubsection{\ackname} The authors acknowledge the financial support by the Federal Ministry of Research, Technology and Space of Germany and by Sächsische Staatsministerium für Wissenschaft, Kultur und Tourismus in the programme Center of Excellence for AI-research „Center for Scalable Data Analytics and Artificial Intelligence Dresden/Leipzig“, project identification number: ScaDS.AI.

\subsubsection{\discintname}
The authors have no competing interests to declare that are relevant to the content of this article.
\end{credits}

%% file: 08-appendix.tex
\section{Appendix}
\label{sec:appendix}

% \begin{table}[h]
% \centering
% \caption{Public agencies included in the dataset (incident reporting and security recommendations).}
% \label{tab:agencies-all}
% \begin{tabular}{@{}cll@{}}
% \toprule
% \textbf{Country} & \textbf{Agency} & \textbf{Abbrev}. \\ \midrule
% AU & Australian Cyber Security Centre & ACSC \\
% AT & Austrian Secure Information Technology Centre & A-SIT \\
% AT & Federal Criminal Police Office & BKA \\

% CA & Canadian Centre for Cyber Security & CCCS \\
% CA & National public awareness campaign ``Get Cyber Safe'' & GCS \\
% CA & Office of the Privacy Commissioner of Canada & OPC \\
% CA & Royal Canadian Mounted Police & RCMP \\

% FI & National Cyber Security Centre & NCSC-FI \\

% FR & Public Action Interest Grouping against Cybermalware & GIP ACYMA \\
% FR & Ministry of Economy and Finance & MEFSIN \\

% DE & Federal Office for Information Security & BSI \\

% JP & NOTICE project (botnet mitigation initiative) & NOTICE \\

% NZ & Ministry of Business, Innovation and Employment & MBIE \\
% NZ & National Cyber Security Centre & NCSC-NZ \\
% NZ & New Zealand Computer Emergency Response Team & CERT NZ \\

% SG & Cyber Security Agency of Singapore & CSA \\
% SG & Singapore Police Force & SPF \\

% UK & City of London Police & CoLP \\
% UK & National Cyber Security Centre & NCSC-UK \\

% US & Cybersecurity and Infrastructure Security Agency & CISA \\
% US & Federal Trade Commission & FTC \\
% US & Internet Crime Complaint Center (FBI) & IC3 \\
% \bottomrule
% \end{tabular}
% \end{table}
\begin{longtable}{@{}cll@{}}
\caption{Public agencies included in the dataset.}
\label{tab:agencies-all} \\
\toprule
\textbf{Country} & \textbf{Agency} & \textbf{Abbrev}. \\
\midrule
\endfirsthead

\toprule
\textbf{Country} & \textbf{Agency} & \textbf{Abbrev}. \\
\midrule
\endhead

\midrule
\multicolumn{3}{r}{\emph{Continued on next page}} \\
\endfoot

\bottomrule
\endlastfoot

AU & Australian Cyber Security Centre & ACSC \\
AT & Austrian Secure Information Technology Centre & A-SIT \\
AT & Federal Criminal Police Office & BKA \\

CA & Canadian Centre for Cyber Security & CCCS \\
CA & National public awareness campaign ``Get Cyber Safe'' & GCS \\
CA & Office of the Privacy Commissioner of Canada & OPC \\
CA & Royal Canadian Mounted Police & RCMP \\

FI & National Cyber Security Centre & NCSC-FI \\

FR & Public Action Interest Grouping against Cybermalware & GIP ACYMA \\
FR & Ministry of Economy and Finance & MEFSIN \\

DE & Federal Office for Information Security & BSI \\

JP & NOTICE project (botnet mitigation initiative) & NOTICE \\

NZ & Ministry of Business, Innovation and Employment & MBIE \\
NZ & National Cyber Security Centre & NCSC-NZ \\
NZ & New Zealand Computer Emergency Response Team & CERT NZ \\

SG & Cyber Security Agency of Singapore & CSA \\
SG & Singapore Police Force & SPF \\

UK & City of London Police & CoLP \\
UK & National Cyber Security Centre & NCSC-UK \\

US & Cybersecurity and Infrastructure Security Agency & CISA \\
US & Federal Trade Commission & FTC \\
US & Internet Crime Complaint Center (FBI) & IC3 \\
\end{longtable}

\newcolumntype{L}[1]{>{\raggedright\arraybackslash}p{#1}}

\begin{longtable}{@{} L{0.4\textwidth} L{0.6\textwidth} @{}}
\caption{Explanations of recommendation labels.}
\label{tab:explained_labels}\\
\toprule
Recommendation label & Explanation \\
\midrule
\endfirsthead

\toprule
Recommendation label & Explanation \\
\midrule
\endhead

\midrule
\multicolumn{2}{r@{}}{\emph{Continued on next page}} \\
\endfoot

\bottomrule
\endlastfoot

Replace after EOL &
Replace routers or smart devices once vendors no longer provide firmware or security updates. \\

Disable remote management &
Disable administrative access to routers or devices from outside the local network (e.g., via the public Internet). \\

Enable automatic updates &
Configure devices or routers to automatically install security and firmware updates. \\

Use guest Wi-Fi &
Place smart home devices on a logically separate network (e.g., guest network or VLAN) to reduce exposure to other household devices. \\

Use WPA2/WPA3 &
Configure the Wi-Fi network to use modern encryption standards (WPA2 or WPA3). \\

Access network over VPN &
Use a virtual private network to securely access the home network remotely rather than exposing management interfaces to the Internet. \\

Check connected devices &
Review the list of devices connected to the home router to identify unknown connections. \\

Monitor internet usage &
Observe network traffic (e.g.\ data volume) that could indicate compromise or misuse. \\

Avoid unnecessary internet &
Limit or disable Internet connectivity for smart devices when external connectivity is not required for their core functionality. \\

Disable unused features &
Turn off device capabilities (e.g., microphones, cameras, Bluetooth, remote access) that are not actively needed, thereby reducing attack surface. \\

Encrypted communication &
Ensure that communication between devices, apps, and cloud services is protected using encryption rather than transmitted in plaintext. \\

Avoid insecure interfaces &
Secure or disable device interfaces and protocols known to lack sufficient security guarantees (e.g., unencrypted web interfaces or legacy protocols). \\

Protect digital identity &
Safeguard personal online accounts and identity information associated with smart home services, including account recovery mechanisms. \\

Use a password manager &
Use dedicated software to generate and store unique, strong passwords across devices and online accounts. \\

Check reviews / ratings &
Evaluate third-party assessments and user reviews before purchasing smart home devices. \\

Limit social media sharing &
Minimize public disclosure of smart home usage, device types, or household routines on social media platforms. \\

\end{longtable}

\begin{longtable}{@{} l l @{}}
\caption{Cybersecurity recommendations mentioned by exactly one governmental agency.}
\label{tab:single_recs_nointerp} \\
\toprule
Agency & Recommendation \\
\midrule
\endfirsthead

\toprule
Agency & Recommendation \\
\midrule
\endhead

\midrule
\multicolumn{2}{r}{\emph{Continued on next page}} \\
\endfoot

\bottomrule
\endlastfoot

ACSC (AU)~\cite{acscWifiRouter} &
Turn off router when on holiday \\
\midrule

\multirow{4}{*}{A-SIT (AT)~\cite{asitSmartHome}} &
Use antivirus software \\
& Child safety \\
& Manage access of smartphone apps to phone \\
& Use VPN to access Internet \\
\midrule

\multirow{2}{*}{A-SIT (AT)~\cite{asitSmartHomeSystems}} &
Use separate router for IoT \\
& Adjust privacy settings in smart device app \\
\midrule

CCCS (CA)~\cite{cccsIotSecurityItsap00012} &
Monitor, detect, and correct security issues \\
\midrule

OPC (CA)~\cite{opcSmartDevicesPrivacy} &
Advise guests about smart devices \\
\midrule

\multirow{2}{*}{NCSC-FI (FI)~\cite{ncscfiHomeNetworkRouter}} &
Reboot router regularly \\
& Check visibility of your IP from the Internet \\
\midrule

\multirow{2}{*}{CISA (US)~\cite{cisaHomeNetworkSecurity}} &
Disable WPS on the router \\
& Reduce Wi-Fi signal strength \\
\midrule

\multirow{2}{*}{IC3 (US)~\cite{ic3Psa250605}} &
Watch out for suspicious behavior \\
& Use only trusted apps \\
\midrule

GIP ACYMA (FR)\cite{cybermalveillanceIot} & Keep phone apps updated

\end{longtable}